\documentclass[twocolumn,showpacs,preprintnumbers,amsmath,amssymb]{revtex4}

\usepackage{graphicx}
\usepackage{dcolumn}
\usepackage{bm}

\begin{document}

\title{Spin Currents in Metallic Nanostructures; Explicit Calculations}

\author{F. S. M. Guimar\~aes$^{(a)}$, A. T. Costa$^{(a)}$, R. B. Muniz$^{(a)}$ and D. L. Mills$^{(b)}$} 
\affiliation{
(a) Instituto de F\'{\i}sica, Universidade Federal Fluminense, Niter\'oi, Brazil \\
(b) Department of Physics and Astonomy; University of California, Irvine; Irvine, California; 92697 U.S.A. }

\date{\today}

\begin{abstract}
In ultrathin ferromagnets deposited on metallic substrates, excitation of precessional motion of the spins produces a spin current in the substrate that transports angular momentum out of the film. This phenomenon is referred to as spin pumping, and is a source of damping of the spin motion. Spin pumping enters importantly in the description of spin dynamics in other nanoscale and subnanoscale systems as well. In this paper, we present an approach based on the Kubo formalism that allows the explicit calculation of this spin current and its spatial variation. We use the formalism to explore features of the spin current generated by spin motions in a simple model system.
\end{abstract}


\maketitle

\section{Introduction}
Spin dynamics in magnetic nanostructures has been under active study for the past two decades. A very important issue is the nature of mechanisms that damp the spin motion. A central question is whether one encounters new processes unique to the nanoscale environment not encountered in bulk matter, where now damping mechanisms are well understood.

The answer to the question just posed is in the affirmative. An extrinsic mechanism referred to as two magnon scattering is evident in magnetic nanostrutures\cite{jindermills}. A second intrinsic mechanism, which is the topic of the present paper, is referred to as spin pumping. For this process to be active the spins of interest must reside in, or be coupled to a metallic structure. The most commonly studied system wherein spin pumping is observed is an ultrathin ferromagnetic film fabricated from a 3d magnetic metal or alloy that is deposited on a metallic substrate. Precession of the spins within the ferromagnetic film generates a spin current in the substrate that transports transverse angular momentum out of the film, in the direction normal to the interface between the film and the substrate. A consequence is that the amplitude of the precessional motion within the film decays with time. 

A theoretical discussion of the consequences of angular momentum transfer between precessing spins and conduction electrons was presented by Slonczewski\cite{slonczewski}, and Berger\cite{berger} explored relaxation of precessing moments by the spin pumping process. In an elegant discussion, \v{S}im\'anek and Heinrich\cite{simanekheinrich} derived a correction to the damping term of the Landau-Lifshitz-Gilbert equation with origin in spin pumping. A short time later one of the present authors\cite{mills2003} and \v{S}im\'anek\cite{simanek2003} provided expanded discussions of spin pumping through use of the approach used in ref. \onlinecite{simanekheinrich}. It should be remarked that the theoretical treatments just cited all employ a simple model wherein the precessing moments are described as localized spins, while the band electrons are treated as free electrons in a parabolic band. These theories thus outline the fundamental physics associated with spin pumping, but their application to real materials is limited. The phenomenon of spin pumping was discovered experimentally by Urban, Woltersdorf and Heinrich\cite{urbanheinrich}. 

In substrates wherein the spin diffusion length is very long, the spin current penetrates deeply. Thus, it can enter a second ferromagnetic film placed downstream to excite its spins through the spin torque mechanism\cite{woltersdorfbeck}. One may envision devices made which exploit this phenomenon, most particularly when spins in the primary film are excited by short pulses. In a recent paper, theoretical studies of pulse excitation of a second magnetic moment through spin current generated by a primary moment have been presented\cite{guimaraesferreira}, in a closely related but somewhat different physical context than ultrathin ferromagnets. 

While the theoretical discussions cited above clearly outline the basic physical origin of the spin pumping phenomenon, as noted they employ a very simple model of the band electrons so their application to real materials is limited. Within the framework of a semi-classical description of spin motions in an ultrathin film sandwiched between two reservoirs, a description of spin pumping applicable to real materials has been developed\cite{tserkovnyakbauer} and calculations that account nicely for data have been presented as well\cite{zwierzyckibauer}. In this approach, one electron theory forms the basis for the discussion. More recently a full quantum theoretic formulation of spin excitations in ultrathin films in actual experimental geometries has been developed and applied to the analysis of ferromagnetic relaxation under circumstances where the spin pumping mechanism is the dominant contribution to the line width\cite{costamills}. Here all features of the problem are described within itinerant electron theory, including the moments whose precession generates the spin current. The discussion is framed within the language of the Kubo formalism of many body theory. Successful accounts of spin pumping data were obtained in parameter free calculations for Fe on Au(100)\cite{urbanheinrich}, and also the theory accounts nicely for systematic features in the line width found in FMR studies of a FM/Cu/FM trilayer deposited on metallic substrates\cite{lenzbaberschke}.

It is ferromagnetic resonance studies of spin waves in ultrathin films that have motivated our earlier theoretical studies of the spin pumping mechanisms just cited. We point out that in ultrathin ferromagnets, spin polarized electron energy loss spectroscopy (SPEELS) can be used to excite and study large wave vector spin waves\cite{vollmerkirschner}. The line width of such modes is so very large that they exist only for two or three cycles of oscillation. The same spin pumping mechanism discussed in the literature on ferromagnetic resonance accounts quantitatively for the line widths found in SPEELS experiments, as discussed in ref. \onlinecite{costamills}.

It should be noted that none of the theoretical studies described above directly explore the spatial dependence of the spin current generated by the precession of the magnetization; as we shall see below, the underlying lattice has strong effects on its spatial structure. The approach in ref. \onlinecite{tserkovnyakbauer} and ref. \onlinecite{zwierzyckibauer} led to formulae only for the flux of spin angular momentum that flows out of the ferromagnetic film, and in ref. \onlinecite{costamills} it is the response function of the spin system that is calculated, with line width extracted through a procedure very similar to that used by experimentalists when data is analyzed. The spin current itself is not calculated in this approach.

It is important to develop the means to calculate the spin current directly, at the microscopic level and to explore its spatial variation, including effects of the underlying lattice. Model calculations presented below show these can be striking. It is noteworthy that temporal dynamics and spatial distribution of pure spin currents have been recently observed by nonlinear optical effects\cite{nonlineartheory,nonlinearexp}. In this paper, we present a discussion within the framework of the Kubo formalism within which such questions can be explored. The approach may be regarded as extension of the methodology of ref. \onlinecite{costamills} and that used in our earlier studies of large wave vector spin waves\cite{costamills2,munizmills}. All electrons in the system, including the moments whose precession is responsible for generation of the spin current, are described within the framework of itinerant electron theory appropriate to the 3d based magnetic materials. Thus we do not resort to a picture within which the precessing spins that generate the spin current are described phenomenologically. So in principle, we are not confined to the one electron description of the materials involved, in contrast to the discussions of ref. \onlinecite{tserkovnyakbauer} and ref. \onlinecite{zwierzyckibauer}. We also present model calculations of the spatial dependence of the spin current in the presence of an underlying lattice, including the influence of a surface on its behavior. This is done for a simple one-dimensional model within which we can examine general aspects of the influence of the underlying lattice on the spin current. We remark that our calculations assume that the electrons propagate through the lattice in a ballistic manner, so our results apply to circumstances where the electron mean free path is long. Room temperature measurements\cite{unguris} of the indirect exchange coupling between Fe films separated by Au layers, all deposited by molecular beam epitaxy (MBE), show excellent agreement with theory based on ballistic electron propagation out to thirty lattice constants, beyond which the exchange cannot be measured. In such high quality MBE grown samples, at low temperature one may expect ballistic descriptions of electron propagation, both in regard to the indirect exchange coupling discussed in ref. \onlinecite{unguris} and the spin pumping currents explored here, to extend to very long distances. We note also quantitative calculations of spin pumping contributions to the linewidth that address real materials, such as those presented in refs. \onlinecite{tserkovnyakbauer,zwierzyckibauer,costamills}, employ ballistic descriptions of electron propagation through the lattice. 

In our view, interest in direct knowledge of the character of the spin current and its spatial distribution is stimulated by observations of the inverse spin Hall effect\cite{saitoh}. If coherent spin motions are excited in an ultrathin film on a metallic substrate, as discussed above, a spin current propagates normal to the interface between the film and the substrate. The spin current contains a DC component. If spin orbit coupling is present, then a DC charge current parallel to the interface is realized. This produces a Hall voltage across the film with magnitude proportional to the square of the amplitude of the spin precession. The ability to perform explicit calculations of the spin current, and its spatial distribution is the first step in a microscopic theory of the inverse spin Hall effect. We note that descriptions of spin dynamics of nanomagnetic systems within which spin orbit coupling is present have appeared very recently\cite{costamills3,khajetoorians}. In the future it should be possible to combine such discussions with analyses of spin currents such as described here to address the inverse spin Hall effect.

The outline of this paper is as follows. Section \ref{theory} is devoted to the theoretical formulation of how one generates microscopic descriptions of spin currents within the framework of the Kubo formalism. Section \ref{calculations} is devoted to presentation of a series of calculations for a model system, and concluding remarks are found in section \ref{conclusion}.

\section{Theoretical Description of Spin Currents in Metals}
\label{theory}

\subsection{General Comments}

It will be useful to provide a bit of orientation regarding the nature of spin currents in metals by recalling the principal results of the analysis found in ref. \onlinecite{tserkovnyakbauer} and ref. \onlinecite{zwierzyckibauer}. These authors assume the magnetic moments in an ultrathin ferromagnet precess coherently and in phase, so the magnetization is characterized by a rigidly rotating vector. We write this as $\vec{M}(t)=M_S\hat{n}(t)$ with $M_S$ the saturation magnetization and $\hat{n}(t)$ a unit vector\cite{versor}. The authors of ref. \onlinecite{tserkovnyakbauer} then show that the spin current radiated into the two semi infinite substrates which surround a ferromagnetic film has the form
\begin{equation}
\label{ipump}
\begin{split}
\vec{I}_S &= \frac{1}{4\pi}\left[A_r\hat{n}(t)\times\frac{d\hat{n}(t)}{dt}-A_i\frac{d\hat{n}(t)}{dt}\right]\\
&= \frac{1}{4\pi M_S^2}\left[A_r\vec{M}(t)\times\frac{d\vec{M}(t)}{dt}-A_iM_S\frac{d\vec{M}(t)}{dt}\right] \ ,
\end{split}
\end{equation}
where we have taken $\hbar = 1$.

The expression in Eq. \ref{ipump} involves vector quantities, whereas a complete description of the spin current requires a tensor object, with one index referring to the direction of transport and the second the spin polarization in the current. In the geometry considered, the direction of transport is normal to the surfaces of the film, and this is not noted explicitly. The vector character thus refers to the spin polarization carried by the current. The expressions for $A_r$ and $A_i$ will be discussed below.

Suppose we consider excitation of the spins by a driving field of frequency $\omega$, as in a ferromagnetic resonance experiment. Then in response to the field, we shall have $\vec{M}(t) = M_S\hat{z}+\delta\vec{M}(t)$, where in the linear response regime $\delta\vec{M}(t)$ oscillates with frequency $\omega$. We then have two components to the spin current. One is an AC component with the frequency $\omega$
\begin{equation}
\label{ipumpac}
\vec{I}_S^{(AC)} = \frac{1}{4\pi M_S}\left[A_r\hat{z}\times\frac{d\delta\vec{M}(t)}{dt}-A_i\frac{d\delta\vec{M}(t)}{dt}\right] 
\end{equation}
and then there is a DC component proportional in strength to the applied microwave power for small amplitude spin motions. This is
\begin{equation}
\label{ipumpdc}
\vec{I}_S^{(DC)} = \frac{A_r}{4\pi M_S}\left\langle \delta\vec{M}(t)\times\frac{d\delta\vec{M}(t)}{dt}\right\rangle\ ,
\end{equation}
where the angular brackets denote a time average. It is the DC component that leads to the inverse spin Hall voltage discussed in the literature, when spin orbit coupling is present in the system. Notice, incidentally, that an AC Hall voltage also is necessarily present.

In our analysis, within the framework of the Kubo formalism, we generate an expression for the spin current. Our approach yields the AC component that, in the formalism of ref. \onlinecite{tserkovnyakbauer}, is described by the expression in Eq. \ref{ipumpac}. However, when an external driving field is applied to the moment or moments responsible for generating the spin current, we calculate $\vec{I}_S^{(AC)}$ directly without the need to resort to explicit descriptions of the one electron eigenstates of the system, and we calculate $\delta\vec{M}(t)$ as well within the same theoretical structure. Thus, if desired we may extract the coefficients $A_r$ and $A_i$ by comparison of our results with Eq. \ref{ipumpac}, which shows the polarization of the AC spin current is perpendicular to the static magnetization. The coefficient $A_i$ controls the magnitude of the component parallel to the time derivative of $\delta\vec{M}(t)$ while $A_r$ controls the component perpendicular to this vector. With $A_r$ in hand, we may use Eq. \ref{ipumpdc} to generate an expression for the DC spin current. 

The point of these remarks is as follows. Our approach is based on linear response theory, and it generates the AC component of the spin current that is first order in the amplitude of the transverse magnetization, whereas the DC component is proportional to its square. Nonetheless if one accepts Eq. \ref{ipump} as a description of the spin current, we can describe the DC component as well in our formalism.

\subsection{A Kubo Formalism Description of the Spin Current Generated by Application of an Exciting Microwave Field to a Moment Bearing Structure}

We now turn to a description of our method. In this paper, we confine our attention to a simple one dimensional lattice of sites, with one orbital per site. It is straightforward to extend what follows to a three dimensional lattice with several orbitals per site by adding appropriate indices, but here we confine our attention to the simple system just described, in the interest of simplicity.

The Hamiltonian that forms the basis of our treatment describes a lattice of sites occupied by atoms that need not be identical:
\begin{equation}
\label{hamilt}
\begin{split}
\hat{H} =& \sum_{\substack{i,j,\sigma\\i\not=j}}t_{ij} \hat{c}_{i\sigma}^{\dag}\hat{c}_{j\sigma}+\sum_{i,\sigma}\epsilon_i\hat{n}_{i\sigma}+\frac{1}{2}\sum_{i,\sigma}U_i\hat{n}_{i\sigma}\hat{n}_{i\overline{\sigma}}\\
&+g\mu_B B_0\sum_i\hat{S}_i^z\ .
\end{split}
\end{equation}
One may view the Hamiltonian in Eq. \ref{hamilt} as describing a “Hubbard alloy”. In Eq. \ref{hamilt}, $t_{ij}$ is the hopping integral between sites $i$ and $j$, $\hat{c}_{i\sigma}^{\dag}$ ($\hat{c}_{i\sigma}$) create (destroy) an electron with spin $\sigma$ at site $i$, $\hat{n}_{i\sigma}=\hat{c}_{i\sigma}^{\dag}\hat{c}_{i\sigma}$, $\epsilon_i$ is the energy of the orbital at site $i$ and $U_i$ represents the on site Coulomb interaction at site $i$. While our formal discussion allows $U_i$ to be non-zero at every site in our system, in our numerical studies we shall set $U_i$ to zero everywhere save for the site $i=N$ where the magnetic moment is located whose precession generates the spin current we study. It is $U_i$ which drives the formation of the magnetic moment at site $N$. Finally $B_0$ is an externally applied magnetic field in the $z$ direction, $\hat{S}_i^z=\frac{1}{2}(\hat{n}_{i\uparrow}-\hat{n}_{i\downarrow})$. We refer to the magnetic moment at site $N$ as $\vec{m}_0$.

We introduce the generalized spin operator $\hat{S}_{ij}^{\mu} = \frac{1}{2}\sum_{\alpha,\beta}\hat{c}_{i\alpha}^{\dag}\sigma^{\mu}_{\alpha\beta}\hat{c}_{j\beta}$ whose diagonal component $\hat{S}_{ii}^{\mu}$ is the $\mu^{th}$ Cartesian component of spin in unit cell $i$. One finds
\begin{equation}
\frac{d\hat{S}_{ii}^{\mu}}{dt}+i\sum_k\left[t_{ki}\hat{S}_{ki}^{\mu}-t_{ik}\hat{S}_{ik}^{\mu}\right]\equiv \frac{d\hat{S}_{ii}^{\mu}}{dt}+\hat{I}_i^{\mu} = 0\ ,
\end{equation}
where $\hat{I}_i^{\mu}$ is the spin current radiated by precessing magnetization in unit cell $i$. The contribution from the terms with $k>i$ describe spin current propagating to the right, and the contribution from terms with $k<i$ describe spin current flowing to the left. If we consider a particular volume $V$ that contains many spins, then the total spin current flowing out of $V$ is
\begin{equation}
\label{isoperator}
\hat{I}_S^{\mu}=\sum_{i\in V}\hat{I}_i^{\mu}=i\sum_{i\in V}\sum_k\left[t_{ki}\hat{S}_{ki}^{\mu}-t_{ik}\hat{S}_{ik}^{\mu}\right]\ .
\end{equation}
In Eq. \ref{isoperator} the sum over $k$ may be restricted to $k\not\in V$ because the contribution to $\hat{I}_S^{\mu}$ from the terms associated with $k\in V$ vanishes. It is useful to define the spin current operator $\hat{I}_S^+=\hat{I}_S^x+i\hat{I}_S^y$, from which one may easily obtain $\hat{I}_S^x$ and $\hat{I}_S^y$. The amplitude of the total spin current pumped into the non-magnetic substrate due to the precession of $\vec{m}_0$ is given by
\begin{equation}
|\langle \hat{I}_S^+\rangle| = \sqrt{|\langle \hat{I}_S^x\rangle|^2+|\langle \hat{I}_S^y\rangle|^2}\ ,
\end{equation}
where $\langle\hat{O}\rangle$ represents the expectation value of the operator $\hat{O}$. It is this quantity we study numerically in the discussion that follows.

The applied transverse magnetic field $\vec{b}_\bot(t)$ couples to the spin density of the magnetic unit located at site $0$, and we shall calculate the change in the expectation value of the spin current caused by this relatively weak time dependent perturbation. Here, the change $\delta\langle I_S^+\rangle =\langle I_S^+\rangle$, because there is no net spin current flowing in the unperturbed system. The interaction with a single frequency oscillatory $\vec{b}_\bot(t)$ may be written as
\begin{equation}
\begin{split}
\hat{H}_I = g\mu_B\vec{b}_\bot(t)\cdot\hat{\vec{S}}_0 &= g\mu_B b_0\left[\cos(\omega t)\hat{S}_0^x-\sin(\omega t)\hat{S}_0^y\right]\\
&= \frac{g\mu_B b_0}{2}\left(e^{i\omega t}\hat{S}_0^+ + e^{-i\omega t}\hat{S}_0^-\right)\ ,
\end{split}
\end{equation}
where the transverse field amplitude $b_0$ is much smaller than $B_0$.

Following the usual steps employed in linear response theory one finds that the spin current generated by this perturbation is given by
\begin{equation}
\label{islinearresponse}
\langle I_S^+ \rangle (t) = \frac{ig\mu_Bb_0}{2} \sum_{\substack{i \in V \\ j \not\in V}} e^{-i\omega t}\left\{ t_{ij}\chi_{ij00}^{+-}(\omega) - t_{ji} \chi_{ji00}^{+-}(\omega) \right\} \ ,
\end{equation}
where
\begin{equation}
\chi_{ijkl}^{+-}(\omega) = \int_{-\infty}^{\infty}dt\ e^{i\omega t }\chi_{ijkl}^{+-}(t)
\end{equation}
is the Fourier transform of the generalized susceptibility
\begin{equation}
\chi_{ijkl}^{+-}(t) = \langle \langle \hat{S}_{ij}^+(t);\hat{S}_{kl}^-(0) \rangle \rangle = - i \Theta(t)\langle[\hat{S}_{ij}^+(t),\hat{S}_{kl}^-(0)]\rangle\ .
\end{equation}
Eq. \ref{islinearresponse} is the central result of this section. It relates the spin current pumped into the non-magnetic substrate to generalized transverse dynamical spin susceptibilities of the system, and we shall proceed next by briefly outlining how these susceptibilities may be determined.

\subsection{Generalized Transverse Spin Susceptibility}

By generating the equation of motion for $\chi_{ijkl}^{+-}(t)$ within the random phase approximation (RPA) one finds 
\begin{equation}
\label{relsusc}
\chi_{ijkl}^{+-}(\omega) = \chi_{ijkl}^0(\omega) - \sum_m\chi_{ijmm}^0(\omega)U_m\chi_{mmkl}^{+-}(\omega)\ ,
\end{equation}
where $\chi_{ijkl}^0(\omega)$ is the noninteracting susceptibility, which may be written as
\begin{equation}
\label{chi0}
\begin{split}
\chi_{ijkl}^{0}(\omega) =& \frac{i}{2\pi} \int_{-\infty}^{\infty}d\omega^{\prime}f(\omega^{\prime})\{[ g_{li}^{\uparrow}(\omega^{\prime}) - g_{li}^{- \uparrow}(\omega^{\prime})] g_{jk}^{\downarrow}(\omega^{\prime} + \omega)\\
&+ [ g_{jk}^{\downarrow}(\omega^{\prime}) - g_{jk}^{- \downarrow}(\omega^{\prime})] g_{li}^{- \uparrow}(\omega^{\prime} - \omega)\}\ .
\end{split}
\end{equation}
Here, $g_{ij}^{\sigma}(\omega)$ and $ g_{ij}^{- \sigma}(\omega)$ represent the time Fourier transforms of the retarded and advanced single-particle propagators for an electron with spin $\sigma$ between sites $i$ and $j$, respectively, and $f(\omega)$ is the Fermi-Dirac distribution function. Direct numerical evaluation of the $\omega^{\prime}$ integrals in Eq. \ref {chi0} along the real axis may be difficult to be realized accurately because the one-electron propagators $g$ and $g^{-}$ usually are not smooth functions of $\omega^{\prime}$. It is possible, however, to extend some those integrals to the complex plane and avoid a great deal of numerical difficulties by rewriting Eq. \ref {chi0} as a sum of three terms ($\chi^{0} = I_1 + I_2 + I_3)$ which are given at zero temperature by \cite{munizmills2}:
\begin{equation}
\label{i1i}
I_1 (\omega)= \frac{1}{2\pi} \int_{\eta}^{\infty}dy g_{li}^{\uparrow}(\omega_F + iy) g_{jk}^{\downarrow}(\omega_F + iy + \omega)
\end{equation}
\begin{equation}
\label{i2i}
I_2 (\omega) = \frac{1}{2\pi} \int_{\eta}^{\infty}dy \big{[} g_{kj}^{\downarrow}(\omega_F + iy) g_{il}^{\uparrow}(\omega_F + iy - \omega)\big{]}^{\ast}
\end{equation}
\begin{equation} \label{i3i}
I_3 (\omega) = -\frac{i}{2\pi} \int_{\omega_F-\omega}^{\omega_F}d\omega^{\prime} g_{li}^{- \uparrow}(\omega^{\prime}) g_{jk}^{\downarrow}(\omega^{\prime} + \omega) \hspace{0.2cm},
\end{equation}

Once all the elements of $\chi^0(\omega)$ have been determined, Eq. \ref{relsusc} may be solved in matrix form and one may finally obtain the interacting susceptibility 
\begin{equation}
\label{relhfrpa}
\hat{\chi}^{+-}(\omega) = [\hat{1} + \hat{\chi}^0(\omega)\hat{U}]^{-1}\hat{\chi}^{0}(\omega) \hspace{0.2cm},
\end{equation}
where $\hat{U}_{mn} = U_m\delta_{mn}$. 

\section{Numerical Studies of the Spin Pumping Current}
\label{calculations}

In this section, we present the results of our studies of the spatial variation of the spin pumping current. We shall also compare the results obtained with our approach with those generated by the method described in ref. \onlinecite{tserkovnyakbauer} and ref. \onlinecite{zwierzyckibauer}. We have used the simple one dimensional model from section \ref{theory} for this first set of calculations.

We take nearest neighbor hopping only into account, and we set the hopping integral $t=1$, so the bandwidth is 4 units. The zero of the energy scale is the middle of the band, so if the band is half filled the Fermi energy $E_F=0$. We set the Coulomb interaction $U_i$ to zero everywhere save for the magnetic moment bearing site $0$ where we take $U_0=10$. We also set the orbital energies $\epsilon_i$ to zero everywhere save for site $0$ and $\epsilon_0$ is adjusted so the occupancy $n_0=1$.

\begin{figure}
\includegraphics[width = 8.cm] {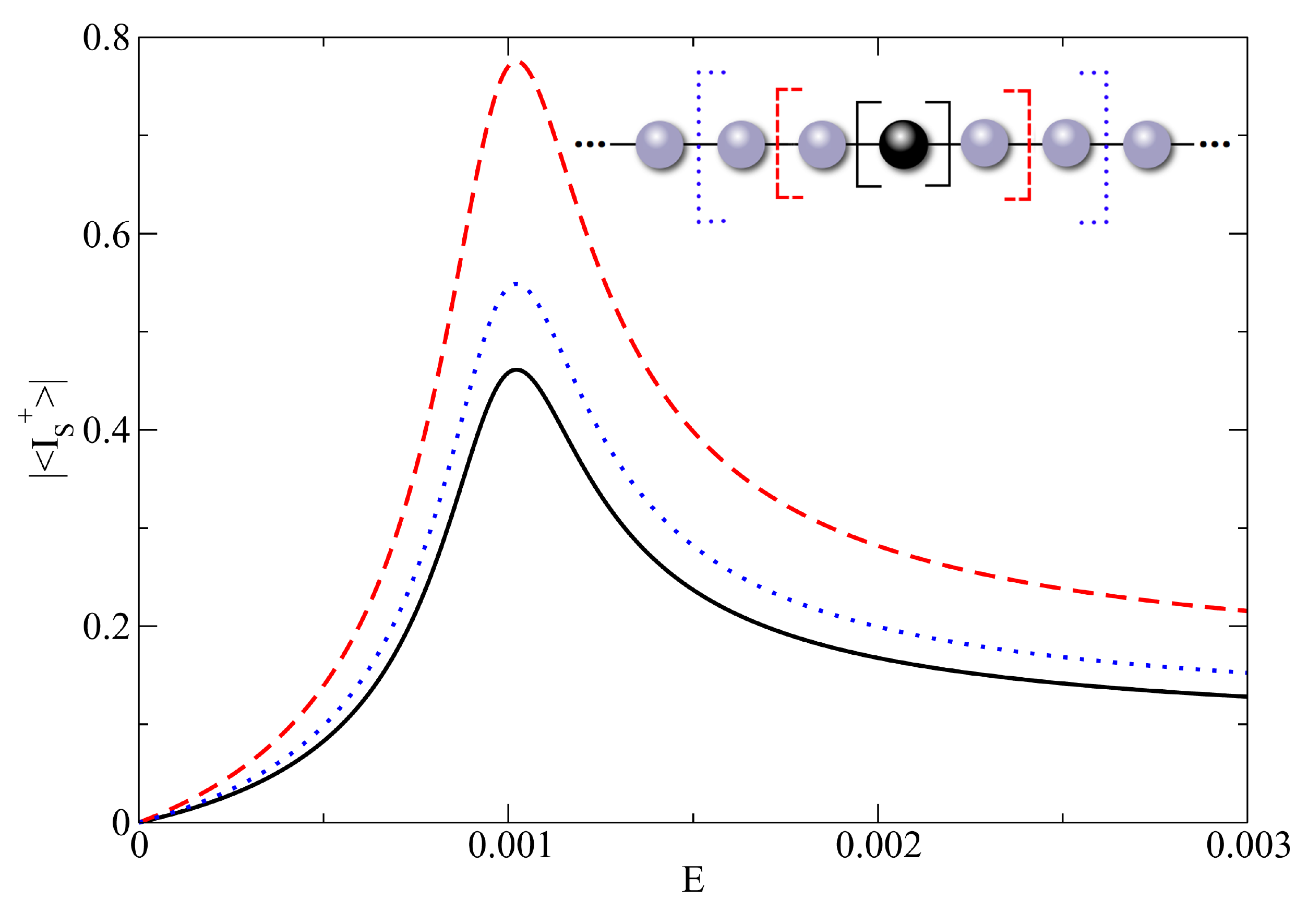}
\caption{Modulus of the expectation value of the spin current $|\langle \hat{I}_S^+\rangle|$ produced by the precession of the moment at site $0$, calculated as a function of the energy $E=\hbar\omega$ across the boundaries shown in the inset. The spin current is in units of $g\mu_B b_0$ and the Zeeman field is adjusted so the resonance frequency of the impurity is $E_0=\hbar\omega_0=10^{-3}$. These calculations are for the case where the band is half filled. }
\label{figure_1}
\end{figure}

We begin by considering the moment embedded in a line of infinite length. The parameters mentioned above give a self-consistent moment on site $0$ of $0.916\mu_B$. In this case, spin current is emitted by the precessing impurity both to the left and to the right. In Fig. \ref{figure_1}, for the case where the band is half filled, we show the modulus of the expectation value of the total spin current, $|\langle \hat{I}_S^+\rangle|$, as a function of energy $E=\hbar\omega$, where $\omega$ is the frequency of the driving field, calculated across different boundaries shown in the inset. The Zeeman field has been adjusted so that the resonance frequency of the impurity is $E_0=\hbar\omega_0 = 10^{-3}$. We see a resonance centered very close to $E_0$. Clearly there is spatial variation in the spin current, and also it is evident that the moment emits an appreciable spin current even for precession frequencies well above $\omega_0$. We remark that, as we see from Fig. \ref{figure_1}, our Kubo formalism based approach allows us to directly calculate the absolute  amplitude of the spin current as a function of the frequency that drives the spin system, with magnitude of the external driving field fixed. We do not need to link our analysis into the Landau-Lifshitz-Gilbert equation as in earlier studies\cite{simanekheinrich,mills2003,simanek2003}.

\begin{figure}
\includegraphics[width = 8.cm] {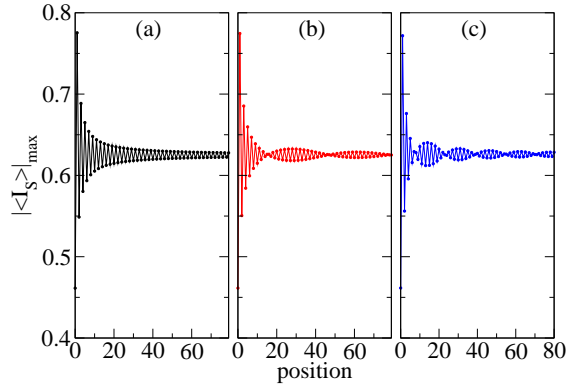}
\caption{We show the spatial dependence of the spin current for three choices of the Fermi energy, (a) $E_F=0$ (band half filled), (b) $E_F=0.1$, and (c) $E_F=0.2$. The impurity is driven on resonance in these calculations. }
\label{figure_2}
\end{figure}

In Fig. \ref{figure_2}, we show the spatial variation of the spin current, for the case where the impurity is excited on resonance. In Fig. \ref{figure_2}(a), the calculation is for the case where the band is half filled ($E_F=0$). The magnitude of the spin current decays as one moves from the impurity, to approach a constant value at large distances. We see oscillations at half the Fermi wavelength about the average value of the spin current. When the band is half filled, the period of oscillations $\Lambda=\frac{\pi}{k_F}$ is twice the lattice constant. In Fig. \ref{figure_2}(b) and Fig. \ref{figure_2}(c) we show the spatial variation for the case where the Fermi energy is moved off half filled. The Fermi wave vector for these two cases is incommensurate with the lattice, and we see the long wavelength oscillations that are very similar in nature to the phenomenon of “aliasing” that has entered earlier discussions of coupling between ferromagnetic films mediated by a metallic non magnetic spacing layer\cite{edwardsphan}. We expect the phenomenon of ``dynamic aliasing'' illustrated in Fig. \ref{figure_2} will survive in calculations of the spin current generated by the coherent precession of spins in a film adsorbed on a substrate, since it is evident in calculations of the static interfilm couplings within the framework of three dimensional analyses\cite{edwardsphan}. 

It is of interest to compare the results of our calculations with the picture set forth in ref. \onlinecite{tserkovnyakbauer} and ref. \onlinecite{zwierzyckibauer}. For the AC component of the spin current, Eq. \ref{ipumpac} provides the relation
\begin{equation}
I_{S}^+ = \frac{i}{4\pi M_S}\left(A_r+iA_i\right)\frac{d\delta M^{+}}{dt}\ .
\end{equation}

The amplitude of the spin motion may be generated from the Landau-Lifshitz-Gilbert equation that is modified in form in the presence of spin pumping. The authors of ref. \onlinecite{tserkovnyakbauer} and ref. \onlinecite{zwierzyckibauer} obtain the form
\begin{equation}
\label{llg}
\frac{d\vec{M}}{dt} = \gamma'\left[\vec{M}\times\vec{B}(t)\right] + \frac{\alpha'}{M_S}\left[\vec{M}\times\frac{d\vec{M}}{dt}\right]\ ,
\end{equation}
where $\gamma^{\prime} = \frac{4\pi m_0}{4\pi m_0+ g A_i} \gamma$ and $\alpha^{\prime} = \frac{4\pi m_0\alpha+g A_r}{4\pi m_0+ g A_i}$. Here $\gamma=-g\mu_B$ is the gyromagnetic ratio, $m_0$ is the magnetic moment of the impurity in Bohr magnetons, and $\alpha$ describes the intrinsic damping rate when spin pumping is absent. In our comparison between the two sets of calculations, we set $\alpha$ to zero. The magnetization is exposed to the magnetic field $\vec{B}(t)=B_0\hat{z}+\vec{b}_\bot(t)$, where $\vec{b}_\bot(t) = b_0\left[\cos(\omega t)\hat{x}-\sin(\omega t)\hat{y}\right]$. One finds, after linearizing Eq. \ref{llg}
\begin{equation}
\label{ipumpplus}
I_{S}^+ = - g\mu_B b_0\left[\frac{A_r+iA_i}{4\pi m_0+ gA_i}\right]\left[\frac{m_0\omega}{(\gamma'B_0-\omega) + i\alpha'\omega}\right] e^{-i\omega t} .
\end{equation}
To proceed with the comparison, one needs to link the coefficients $A_r$ and $A_i$ to our formalism. From ref. \onlinecite{tserkovnyakbauer}, which employs a ballistic description of electron propagation as we do here, one has
\begin{subequations}
\label{arai}
\begin{equation}
A_r = \frac{1}{2}\sum_{mn}\{|r_{mn}^{\uparrow} - r_{mn}^{\downarrow}|^2+|t_{mn}^{\uparrow} - t_{mn}^{\downarrow}|^2\}
\end{equation}
and
\begin{equation}
A_i = \textrm{Im}\sum_{mn}\{r_{mn}^{\uparrow}(r_{mn}^{\downarrow})^{\ast}+t_{mn}^{\uparrow}(t_{mn}^{\downarrow})^{\ast}\}\ .
\end{equation}
\end{subequations}
In these expressions, $r_{mn}^{\sigma}$ and $t_{mn}^{\sigma}$ are reflection and transmission coefficients for an electron with spin $\sigma$ at the Fermi energy $E_F$ scattered by the magnetic unit from channel $m$ to channel $n$ of the normal metal. For our one dimensional model, there is only one channel available to both the initial state and the final state, so $r_{mn}^{\sigma}=r^\sigma\delta_{mn}\delta_{n0}$ and similarly for $t_{mn}^{\sigma}$. One finds\cite{economou} $r^\sigma=\frac{V^\sigma g^0_{00}(E_F)}{1-V^\sigma g^0_{00}(E_F)}$ and $t^\sigma=\frac{1}{1-V^\sigma g^0_{00}(E_F)}$, where $V^{\sigma}=-\sigma\frac{Um}{2}$ with $\sigma=\pm1$ for majority ($\uparrow$) and minority ($\downarrow$) spins, respectively, and $g_{00}^0$ is the unperturbed one-electron propagator at site $0$. 

\begin{figure}
\includegraphics[width = 8.cm] {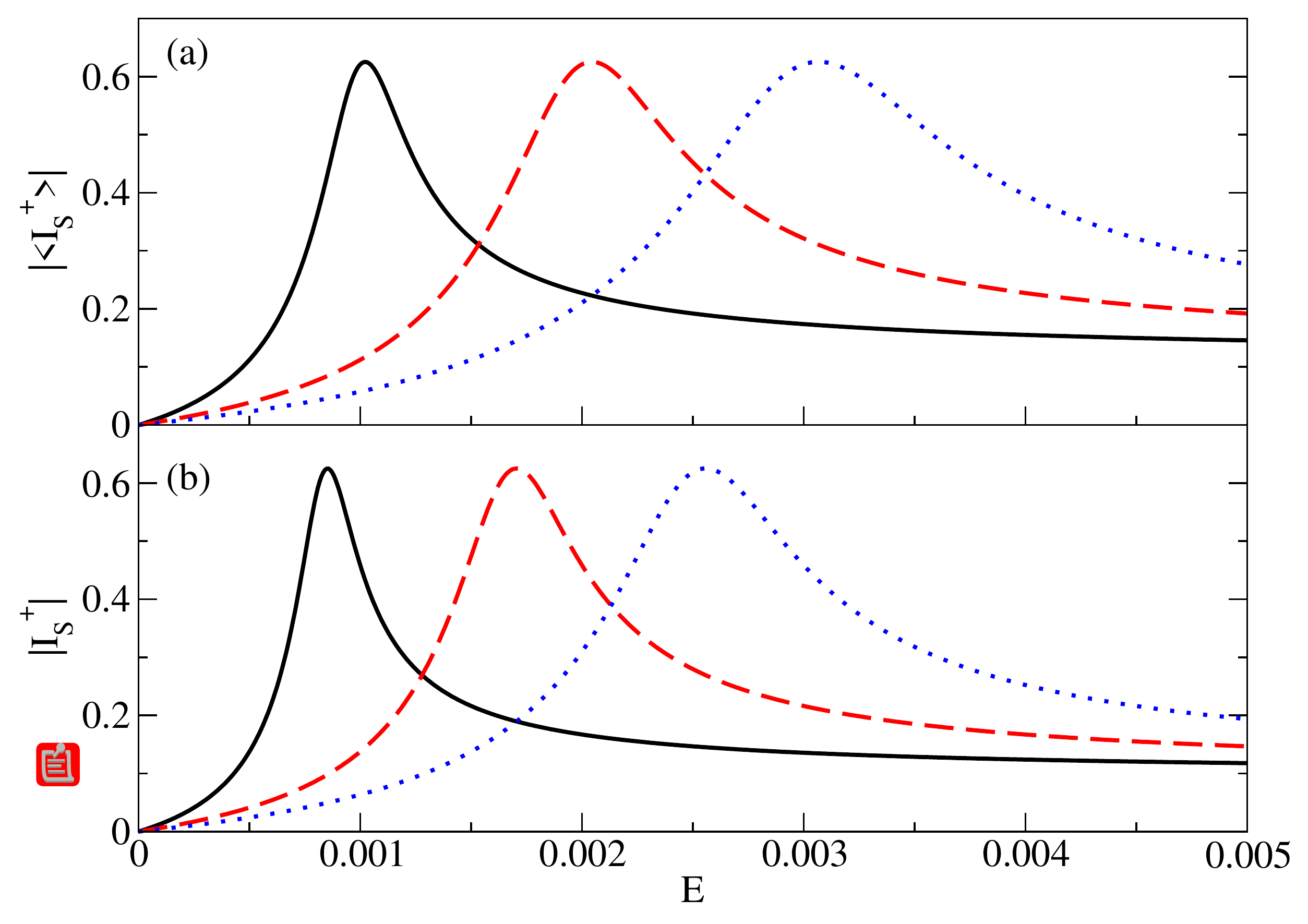}
\caption{We show a comparison between (a) the spin current far from the precessing impurity calculated with our formalism, and (b) the spin current as described by the formalism of ref. \onlinecite{tserkovnyakbauer} and ref. \onlinecite{zwierzyckibauer}. The solid, dashed and dotted lines refer to calculations for which the Zeeman energy $E_0=\hbar\omega_0$ is taken to be $1.0\times10^{-3}$, $2.0\times10^{-3}$ and $3.0\times10^{-3}$, respectively. }
\label{figure_3}
\end{figure}

In Fig. \ref{figure_3}, we present a comparison between the results of our formalism, and the predictions that follow when Eq. \ref{ipumpplus} is used in conjunction with Eqs. \ref{arai}. Both sets of calculations have been performed for identical model parameters. In Fig. \ref{figure_3}(a) we show the results generated with the formalism developed in the present paper. Here we evaluate the spin current very far away from the impurity, where the spin current has settled down to its constant, asymptotic value. Fig. \ref{figure_3}(b) gives the results from the formalism of ref. \onlinecite{tserkovnyakbauer}. It is the case that the calculations based on our formalism give a renormalized resonance frequency slightly higher than that which follows from ref. \onlinecite{tserkovnyakbauer}. We suggest this has its origin in the fully dynamical nature of our analysis, wherein the spin current can react back onto the local moment and affect its precession frequency. We noted a similar effect earlier when full dynamical calculations of spin wave dispersion relations in ultrathin films are compared by those generated by adiabatic theory\cite{costamills2}. In the present case, the difference is rather small.

If the precessing impurity is near a surface, the spatial variation of the spin current radiated from it differs dramatically from the case where it is in the bulk of an infinitely extended material. In the latter case, spin current streams both to the right and the left, and symmetry requires the amplitude of each stream to be the same. However, as argued some time ago in a phenomenological discussion presented by Valet and Fert\cite{valetfert}, the spin current must vanish at the surface. If the surface is located at $z=L$, then between our impurity and the surface, or in a system with a film radiating spin current into a substrate of finite thickness, in Valet-Fert theory the spatial profile of the spin current will have the form $\sinh\left(\frac{L-z}{\lambda_S}\right)$ where $\lambda_S$ is the spin diffusion length. In our model, the spin diffusion length is infinite, so within the phenomenological theory the spin current should have a linear variation, proportional to $(L-z)$.

An alternative explanation for the linear variation of the spin current is the following: at an arbitrary position $z$ between the magnetic impurity and the surface, the total spin current $I$ is given by the difference between the current $I_r$ flowing toward $z = L$ and the current $I_\ell$ flowing toward the impurity located at $z = 0$. Both $I_r$ and $I_\ell$ are proportional to the direction of the transverse component of the electron spin at $z$, which is given by $\hat{\sigma}_\bot(\tau) = \cos(\omega\tau)\hat{x} + \sin(\omega\tau)\hat{y}$, where $\tau$ is the time the perturbation takes to reach $z$. Since the spin disturbance propagates with the electronic Fermi velocity, $\tau_l = z/v_F$ and $\tau_r = (2L-z)/v_F$. For relatively small frequencies, it is possible to show that $|I|\propto |2\omega(z-L)/v_F|$.

\begin{figure}
\includegraphics[width = 8.cm] {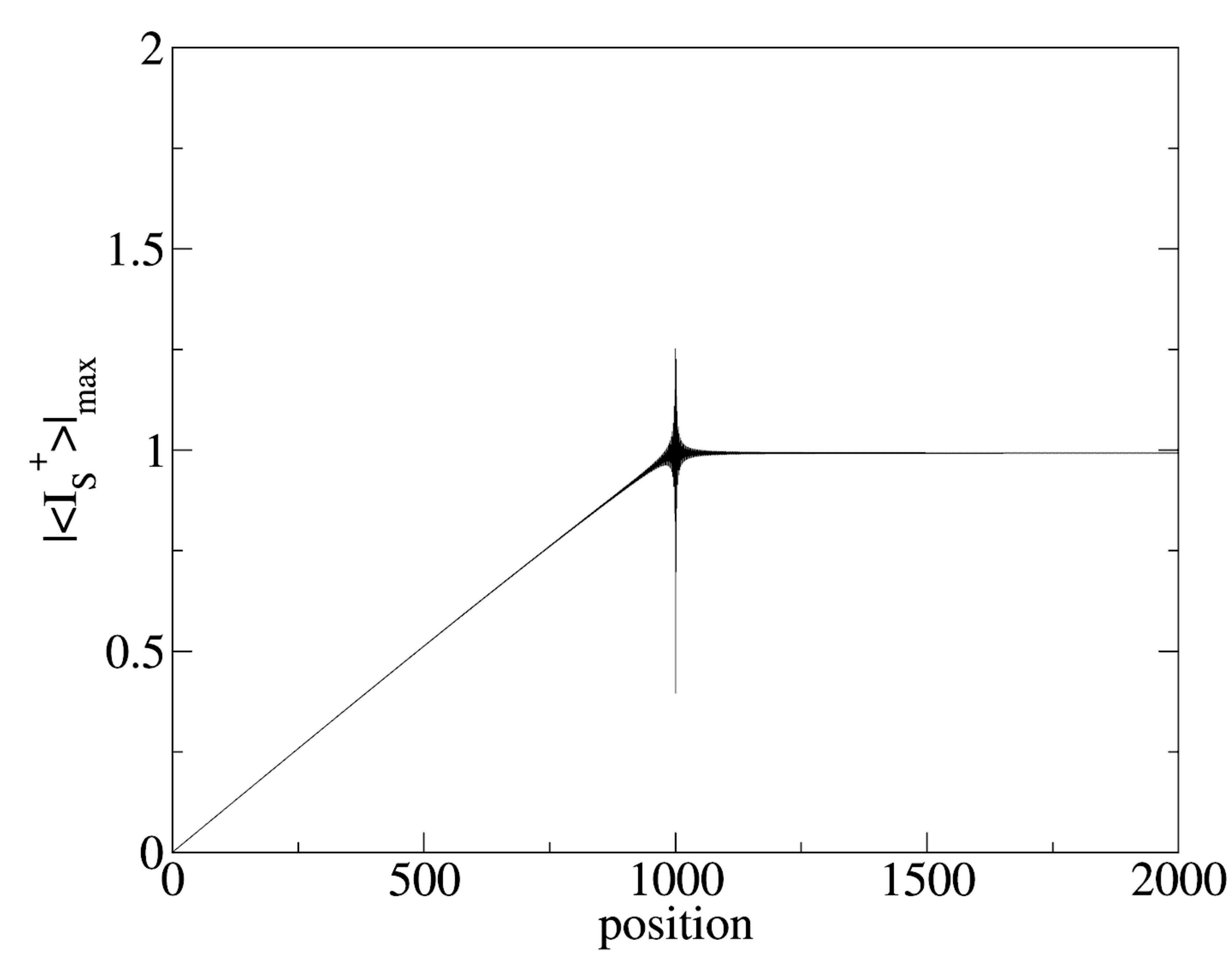}
\caption{The spatial variation of the spin current generated by a magnetic moment which precesses in a semi-infinite line of lattice sites, calculated on resonance. The surface is at $N=1$, and the precessing impurity is located at the site $N=1001$. In this calculation we set $E_F=0$.}
\label{figure_4}
\end{figure}

In Fig. \ref{figure_4}, we show results of the spin current generated by a precessing moment in a semi-infinite line of sites, calculated on resonance. The surface is located at site $N=1$, and the precessing impurity is located at site $N=1001$. We see that to the right of the precessing moment, the spin current settles down to a constant value as angular momentum is carried off into the bulk of the material. However, on the left side, in the region between the moment and the surface, our microscopic calculation produces a linear decrease of the spin current in excellent agreement with the phenomenology set forth in ref. \onlinecite{valetfert}. The surface influences the spatial variation in the spin current even when the precessing moment is very far from the surface. Of course, in a real material with finite spin diffusion length $\lambda_S$ the moment will not feel the presence of the surface when the distance between it and the surface is long compared to $\lambda_S$. In high purity materials such as the noble metals, the spin diffusion length can be quite long, it should be noted. One can grow ultrathin 3d ferromagnets on semiconducting substrates such as GaAs. It would be of great interest to cap such films with a noble metal such as Ag (indeed this is often done to prevent oxidation of the ferromagnet) and then study the dependence of the spin pumping relaxation rate as a function of thickness of the capping layer.

From Fig. \ref{figure_4}, in our model with its infinite spin diffusion length, it is clear that the presence of the surface strongly influences the nature of the spin current radiated from the precessing impurity.
One can then inquire how, as the impurity is moved farther and farther from the surface, one approaches the spin pumping line width realized in the bulk of the material. This can be studied by generating the response function associated with the impurity moment itself. In principle, one can measure the width of the calculated resonant structure as a function of distance of the moment from the surface, to see how the line width approaches the bulk value.

\begin{figure}
\includegraphics[width = 8.cm] {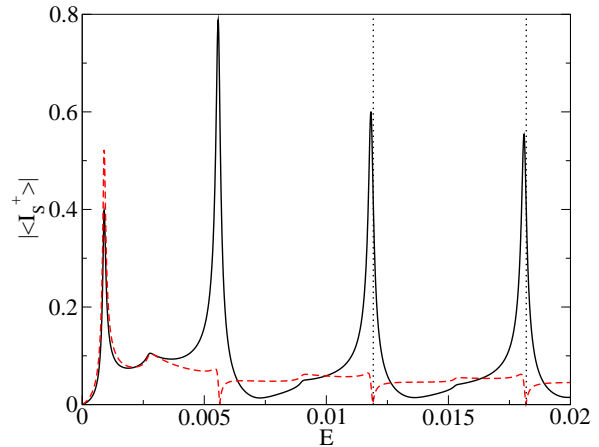}
\caption{The energy variation of the spin current radiated by an impurity located at position $N=1001$. The calculations show the influence of the quantum well states between the impurity position and the surface. Illustrated is the spin current calculated between sites $1001$ and $1000$ (solid line) and the spin current radiated off into the bulk, calculated between sites $1001$ and $1002$ (dashed line). The vertical dotted lines indicate the energies at which the results depicted in Fig. \ref{figure_6} are evaluated.}
\label{figure_5}
\end{figure}

We have studied this issue, to find that the approach to the bulk limit is not simple, for the following reason. When the impurity is at any given position $N$, the region between site $N$ and the surface at $N=1$ acts as a quantum well of finite depth, by virtue of the fact that an electron reflected from the surface is backscattered by the one electron potential associated with the impurity site. There are then resonant levels for electrons in this region, and these produce dramatic structures in the resonance lineshape. Under these circumstances, it is difficult to define the line width. We illustrate such structures in Fig. \ref{figure_5}, where we plot the frequency variation of the spin current in the region between the impurity and the surface (solid line) and also the frequency variation of the spin current radiated into the bulk (dashed line), both calculated across the bond between the magnetic impurity and its nearest neighbors. Similar structures are found in the lineshape of the impurity response as well. As the impurity is moved farther and farther from the surface, these subsidiary peaks move inward to merge with the main structure in the resonant response. It is thus not clear how one can extract something that can be called a line width when the profile of the response function is so complex. It is the case that our calculation produces a bulk like response very far indeed from the surface, if one blurs out the fine structure produced by the resonant states.

\begin{figure}
\includegraphics[width = 8.cm] {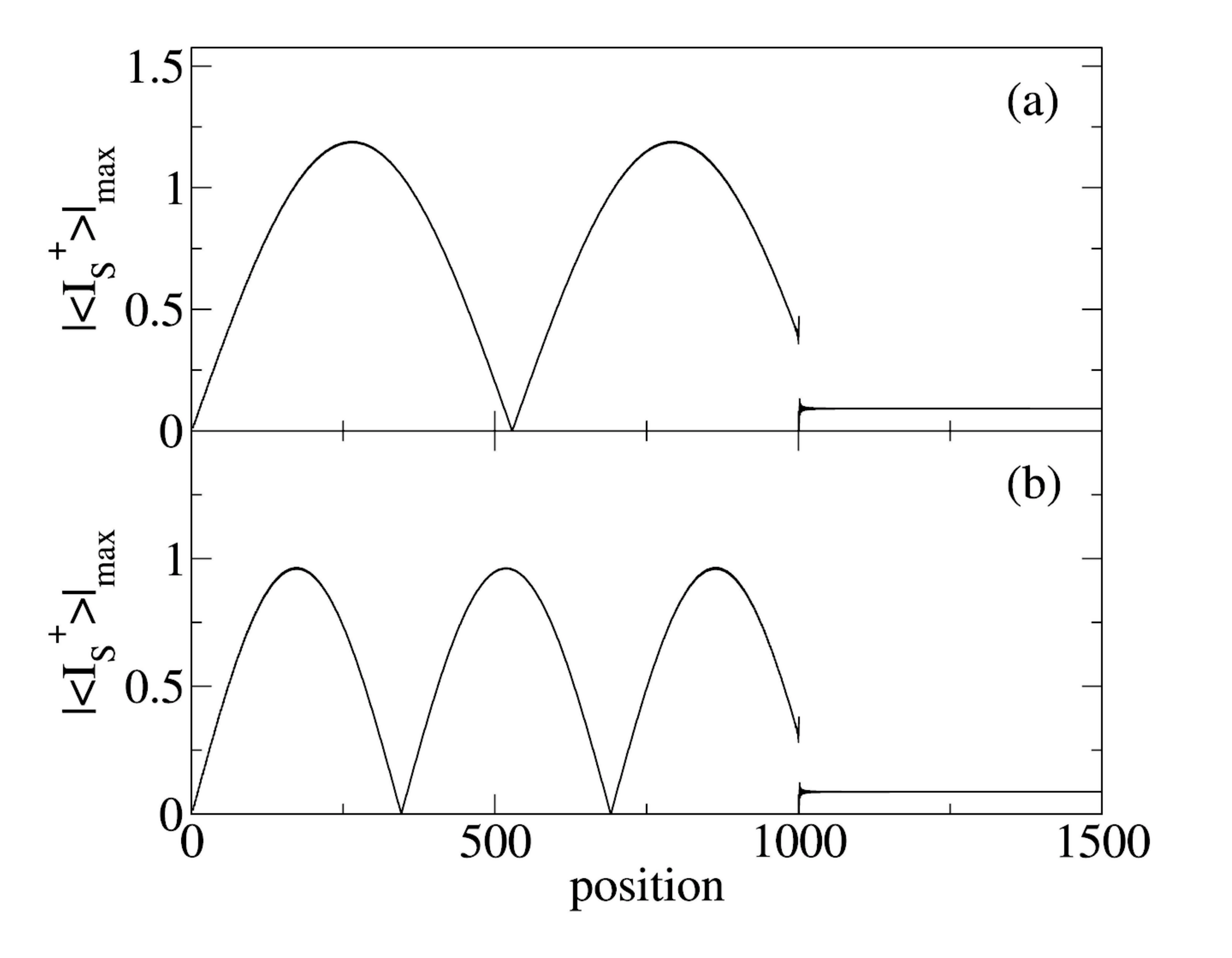}
\caption{The spatial variation of the spin current generated by a magnetic moment which precesses in a semi-infinite line of lattice sites, calculated at energies (a) $E = 0.0119$ and (b) $E = 0.0182$, depicted by vertical dotted lines in Fig. \ref{figure_5}. The surface is at $N=1$, and the precessing impurity is located at the site $N=1001$. In this calculation we set $E_F=0$. }
\label{figure_6}
\end{figure}

The spatial variation of the spin current in the region between the impurity and the surface is most striking, for excitation frequencies in the vicinity of the subsidiary resonances illustrated in Fig. \ref{figure_5}. We illustrate this in Fig. \ref{figure_6}. In Fig. \ref{figure_6}(a), we see that the spatial variation is controlled by a single resonance level (“virtual level”) in the region between the impurity and the surface, and the same is true in Fig. \ref{figure_6}(b). A proper quantum theory is required for such features to appear. They are absent from the macroscopic approach described by Valet and Fert\cite{valetfert}, for example. It will be of great interest to see if such features will be present in a full multiband description of spin pumping relaxation in capped ultrathin films. If the magnetization in such a film is excited by a microwave field, all spins in a given plane parallel to the surface and interface will precess in phase. The physical situation is thus similar to that in the one dimensional model explored in the present paper. We shall explore the question of resonant structures in the relaxation rate for realisitic three dimensional electronic structures in the near future.

It is our view that the features described in the discussion above, while illustrated through use of our one dimensional model, are general phenomena that will be present in full three dimensional calculations directed toward ferromagnetic resonance excitation of coherent, spatially uniform spin motions in an ultrathin film on a substrate or such a film with a metallic capping layer. We note that the existence of quantum well states have been directly observed in magnetic multilayers by inverse angle-resolved photoemission\cite{quantumwell}.

\begin{figure}
\includegraphics[width = 8.cm] {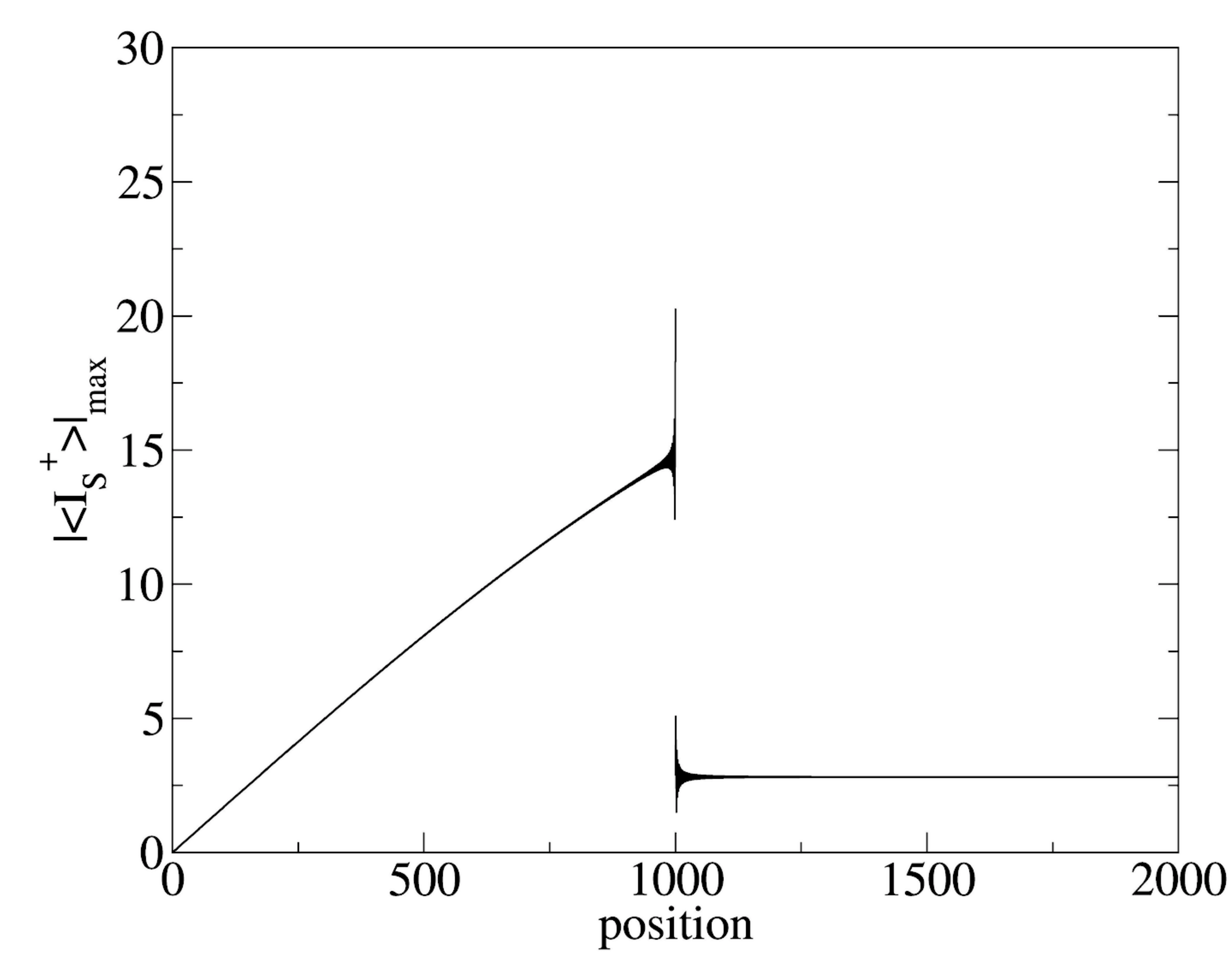}
\caption{The spatial variation of the spin current generated by a magnetic moment which precesses in a semi-infinite line of lattice sites, calculated at resonance. The surface is at $N=1$, and the precessing impurity is located at the site $N=1000$. In this calculation we set $E_F=0$. }
\label{figure_7}
\end{figure}

We now turn to effects specific to our one dimensional model. An example of a physical system wherein these considerations will be relevant is magnetic moments on carbon nanotubes, as discussed in ref. \onlinecite{guimaraesferreira}. There is an even/odd effect in the spatial distribution of spin current, in the presence of a surface. In Fig. \ref{figure_4}, with the surface site $N=1$, we have illustrated the spatial distribution of spin current for an impurity $1000$ sites removed from the surface, at $N=1001$. In Fig. \ref{figure_7}, we show what happens when the impurity is moved one lattice site closer to the surface, $N=1000$. In both figures, the impurity is excited at resonance and the calculations assume the band is precisely half filled. We see in both figures that the spin current vanishes at the surface, as the boundary condition in ref. \onlinecite{valetfert} requires. However, when the impurity is on a site where $N$ is even, there is a discontinuity in the spin current at the impurity itself. This is a quantum effect; the period of oscillations $\Lambda$ for a half filled band is two lattice constants, so movement of the impurity one lattice constant leads to a phase change of 180 degrees for any disturbance it samples from electrons at or very close to the Fermi surface. The spin current near the impurity on the left side of the figure is a coherent superposition of that radiated by the impurity and directed toward the surface, and that reflected back off the surface. There are quantum interference effects associated with these two counter propagating currents, so movement of the impurity by a single lattice constant affects the total spin current samples by it by a large amount. We remind the reader that our calculations are all in the limit where ballistic transport is envisioned. Such interference effects will be absent in a system wherein the electron mean free path is shorter than the distance between the impurity and the surface.

\section{Concluding Remarks}
\label{conclusion}

In this paper we have developed a formalism that allows the calculation of the spatial dependence of the spin current generated by precessing magnetic moments embedded within or in contact with conducting materials. The discussion is carried out entirely within an itinerant electron description appropriate to metallic 3d ferromagnets. The magnetic moments whose precession generate the spin current are formed through action of on site, intra atomic Coulomb interactions of Hubbard character as opposed to the localized spin models used in other approaches. In this paper, in the interest of simplicity we have presented the formalism framed in language suitable to a one dimensional system. The extension to a three dimensional system described by the empirical tight binding model we have used in earlier studies\cite{costamills,costamills2,munizmills} is very straightforward. We have calculations underway directed toward real materials. 

We have also presented a series of numerical calculations of the spatial variation of the spin current generated by a single precessing magnetic moment embedded in a one dimensional lattice of sites, infinite or semi-infinite in length. For the case where the moment is embedded in an infinite line of sites, we find striking spatial variations of the spin current at microscopic length scales that have their origin in the Fermi wavelength of the electrons at the Fermi surface, and its relationship to the underlying lattice constant. These oscillations, as illustrated in Fig. \ref{figure_2}, are a dynamic analogue of the phenomenon of aliasing that enters the description of the static RKKY interaction between ferromagnetic films separated by a metallic film\cite{edwardsphan}. We can expect such features to be present also in three dimensional calculations, it should be remarked. For instance, if one considers an ultrathin ferromagnetic film on a metallic substrate excited in ferromagnetic resonance, in each layer of the ferromagnet the spins precess in phase. In this circumstance, each layer may be viewed as a giant spin embedded in a one dimensional lattice. The spin current in the substrate will also vary only in the direction perpendicular to the interface, and it will be constant throughout a given plane. There is thus an analogy between this circumstance and the one dimensional model system explored in the present paper.

We have also presented studies of the influence of a surface on the spatial distribution of spin current. Our microscopic model produces results in excellent accord with the phenomenological discussion of Valet and Fert\cite{valetfert}. The even/odd effect discussed near the end of Section \ref{calculations} is fascinating to us. While this is specific to one dimensional systems, it is our view that in carbon nanotubes wherein the spin diffusion length is very long, the considerations in our discussion are relevant to the description of the relaxation rate of a localized moment incorporated into such a structure.

\section*{Acknowledgements}

F.S.M.G., A.T.C. and R.B.M. acknowledge support received from CNPq through Instituto do Mil\^enio de Nanotecnologia, Brazil. The research of D.L.M. has been supported by the U. S. Department of Energy, through grant No. DE-FG03-84ER-45083.

\bibliographystyle{abbrv} 

\begin{thebibliography}{99}
\bibitem{versor} Our notation differs from that used in ref. \onlinecite{tserkovnyakbauer} and ref. \onlinecite{zwierzyckibauer}. These authors use $\vec{m}(t)$ in place of our $\hat{n}(t)$.
\end{thebibliography}

\end{document}